\documentclass{ws-ijmpe}
\usepackage[super,compress]{cite}
\usepackage{graphics}
\usepackage{amsmath}
\usepackage{amssymb}
\begin{document}
\markboth{Yash Kaur Singh, R. Chandra, K. Chaturvedi, Tripti Avasthi,
P.K. Raina and P.K. Rath}
{Nuclear transition matrix elements for neutrinoless double-$\beta $
decay within mechanisms
%%involving light Majorana neutrino mass and right-handed current
}

\catchline{}{}{}{}{}
%
%%%%%%%%%%%%%%%%%%%%%%%%%%%%%%%%%%%%%%%%%%%%%%%%%%%%%%%%%%%%%%%%%%%%

\title{Nuclear transition matrix elements for neutrinoless double-$\beta $
decay within mechanisms involving light Majorana neutrino mass and
right-handed current}

\author{Yash Kaur Singh and R. Chandra\footnote{Corresponding Author}}
\address{Department of Physics, Babasaheb Bhimrao Ambedkar
University\\
Lucknow, Uttar Pradesh-226025, India\\
$^*$ramesh.luphy@gmail.com}

\author{K. Chaturvedi}

\address{Department of Physics, Bundelkhand University\\
Jhansi, Uttar Pradesh-284128, India}

\author{Tripti Avasthi and P. K. Rath}

\address{Department of Physics, University of Lucknow\\
Lucknow, Uttar Pradesh-226007, India}

\author{P. K. Raina}

\address{Department of Physics, Indian Institute of Technology\\
Ropar, Rupnagar, Punjab-140001, India }

\maketitle

\begin{history}
\received{Day Month Year}
\revised{Day Month Year}
\end{history}

\begin{abstract}
Employing the projected-Hartree-Fock-Bogoliubov (PHFB) model in conjunction
with four different parametrizations of pairing plus multipolar effective
two body interaction and three different parametrizations of Jastrow short
range correlations, nuclear transition matrix elements for the neutrinoless
double-$\beta $ decay of $^{94,96}$Zr, $^{100}$Mo, $^{110}$Pd, $^{128,130}$%
Te and $^{150}$Nd isotopes are calculated within mechanisms involving light
Majorana neutrino mass and right handed current. Statistically, model
specific uncertainties in sets of twelve nuclear transition matrix elements
are estimated by calculating the averages along with the standard
deviations. For the considered nuclei, \ the most stringent extracted
on-axis limits on the effective light Majorana neutrino mass $<m_{\nu }>$,
the effective weak coupling of right-handed leptonic current with
right-handed hadronic current $<\lambda >$, and the effective weak coupling
of right-handed leptonic current with left-handed hadronic current $<\eta >$
\ from the observed limit on half-life $T_{1/2}^{0\nu }$ of $^{130}$Te
isotope are $0.33$ eV, $4.57\times 10^{-7}$ and $4.72\times 10^{-9}$,
respectively.

\end{abstract}

\ccode{PACS numbers: 23.40.Hc, 21.60.Jz}

%\tableofcontents

\section{Introduction}

Observation of the lepton number $L$ violating neutrinoless double beta ($%
0\nu \beta \beta $) decay is the most pragmatic approach to establish the
Majorana nature of neutrinos. Arguably, the violation of lepton number $L$
conservation and Majorana nature of neutrinos are intimately related \cite%
{sche82}. In 0$\nu \beta \beta $ decay, the neutrino emitted from a nucleon
is to be absorbed by another nucleon implying the existence of Majorana
neutrino with finite mass. Alternatively, the occurrence of 0$\nu \beta
\beta $ decay is also possible with the coexistence of right-handed $V+A$
and left-handed $V-A$ currents. In addition, the smallness of neutrino mass
as explained by see-saw mechanism requires gauge groups with right-handed
current. In several alternative mechanisms based on various gauge
theoretical models beyond the standard model of electroweak unification, the
conservation of lepton number $L$ is violated. Specifically, the exchange of
light and heavy Majorana neutrinos involving left and right handed currents
within the left-right symmetric model (LRSM) is one of such possibilities.

The rate of 0$\nu \beta \beta $ decay is a product of appropriate
phase-space factors, nuclear transition matrix elements (NTMEs) and
parameters of the underlying mechanisms \cite{verg16,dell16}. Recently, the
phase-space factors have been calculated to good accuracy incorporating the
screening correction \cite{koti12,stoi13,stef15}. The extraction of accurate
limits on the parameters of a particular mechanism depends on the
reliability of NTMEs. The evaluation of reliable NTMEs is a challenging
task. A suitable truncation of unmanageable Hilbert space into a manageable
model space with appropriate single-particle energies (SPEs), and effective
two-body interaction is required. In addition, alternative considerations of
the finite size of nucleons (FNS), short range correlations (SRC) and the
effective value of axial vector current coupling constant $g_{A}$ are also
available.

The standard mass mechanism of $0\nu \beta ^{-}\beta ^{-}$ decay has been
extensively studied employing a large number of nuclear models, namely
shell-model approach \cite{caur08,mene09,horo14,brow15,senk16}, QRPA \cite%
{simk99,simk08,simk09,civi09,suho12}, QRPA with isospin restoration \cite%
{simk13}, deformed QRPA \cite{faes12,must13},
projected-Hartree-Fock-Bogoliubov (PHFB) \cite{rath10,rath12,rath13}, energy
density functional (EDF) \cite{rodr10}, covariant density functional theory
(CDFT) \cite{yao15}, and interacting boson model (IBM) \cite{bare13} with
isospin restoration \cite{bare15}. The details about these theoretical
studies have been excellently reviewed over the past years in Refs. %
\citen{suho98,faes99,enge16} and references there in. In spite of the fact
that each model employs different model space, SPEs and two-body residual
interactions, the calculated NTMEs $M^{(0\nu )}$ differ by a factor of 2--3.

Uncertainties in NTMEs for $0\nu \beta ^{-}\beta ^{-}$ decay within
mechanisms involving light Majorana neutrino mass, classical Majorons and
sterile neutrinos have been estimated employing the PHFB approach in
conjunctions with four different parametrizations of effective two-body
interaction, form factors with two different parametrizations and three
different parametrizations of the SRC \cite{rath13}. The uncertainties in
NTMEs for $0\nu \beta ^{-}\beta ^{-}$ decay involving heavy Majorana
neutrino mass \cite{rath12} and new Majoron models \cite{rath16} have also
been investigated. The main objective of the present work is to calculate
sets of twelve NTMEs for the $0^{+}\rightarrow 0^{+}$ transition of $0\nu
\beta ^{-}\beta ^{-}$ decay involving light neutrino mass and right-handed
current by employing sets of four different PHFB wave functions as well as
three different parametrizations of SRC and estimate uncertainties therein.

The detailed theoretical formalism of $0\nu \beta ^{-}\beta ^{-}$ decay
within the mechanisms of LRSM, namely the exchange of light as well as heavy
Majorana neutrino, admixture of $V-A$ and $V+A$ currents, and exchange of
right handed heavy neutrino has been developed in Refs.~%
\citen{doi85,tomo91,doi93}. The theoretical formalism of the standard mass
mechanisms has been extended by including the contribution of induced
currents \cite{simk99}. Including the induced pseudoscalar terms in the
nonrelativistic reduction of right-handed $V+A$ current, the light neutrino
exchange mechanism of $0\nu \beta ^{-}\beta ^{-}$ decay with left and right
handed leptonic and hadronic currents has been investigated in detail \cite%
{stef15}. Presently, the NTMEs are calculated neglecting the induced
pseudoscalar terms in the nonrelativistic reduction of right handed $V+A$
current, which will not apparently change the final conclusions as seen in
Ref.~\citen{stef15}. However, this aspect will be delt in future publication.

In Sec. 2, we present a brief theoretical formalism to study $0\nu \beta
^{-}\beta ^{-}$ decay involving light Majorana neutrino mass and right
handed current. The calculated NTMEs required to study 0$\nu \beta ^{-}\beta
^{-}$ decay of $^{94,96}$Zr, $^{100}$Mo, $^{110}$Pd, $^{128,130}$Te and $%
^{150}$Nd isotopes for the $0^{+}\rightarrow 0^{+}$ transition and the
uncertainties in NTMEs are presented in Sec. 3. Further, the extracted
limits on the effective light Majorana neutrino mass $<m_{\nu }>$, the
effective weak coupling of right-handed leptonic current with right-handed
hadronic current $<\lambda >$, and the effective weak coupling of
right-handed leptonic current with left-handed hadronic current $<\eta >$
from the largest available limits on half-lives of 0$\nu \beta ^{-}\beta
^{-} $ decay $T_{1/2}^{\left( 0\nu \right) }(0^{+}\rightarrow 0^{+})$ are
presented in the same section. Conclusions are given in Sec. 4.

\section{Theoretical Formalism}

The general form of weak interaction Hamiltonian $H_{W}$ is given by 
\begin{equation}
H_{W}=\frac {G}{\sqrt{2}}\left[ j_{L\mu }J_{L}^{\mu \dagger }+\kappa j_{L\mu
}J_{R}^{\mu \dagger }+\eta j_{R\mu }J_{L}^{\mu \dagger }+\lambda j_{R\mu
}J_{R}^{\mu \dagger }\right] +h.c.,  \label{haml}
\end{equation}%
where $j_{L,R}$ and $J_{L,R}$ are left and right handed leptonic and
hadronic currents, respectively. Further, $\kappa ,\,\eta $ and $\lambda $
are the parameters for the admixture of $V-A$ and $V+A$ currents. The second
term in the Eq. (\ref{haml}) is usually neglected as $\kappa $ enters into $%
\beta \beta $ decay amplitude always in the combination $1\pm \kappa $ and
it is expected that $\left\vert \kappa \right\vert \ll 1$.

Using the standard approximations of Ref.~\citen{doi85}, with CP
conservation, the rate for the $0^{+}\rightarrow 0^{+}$ transition of 0$\nu
\beta ^{-}\beta ^{-}$ decay is given by 
\begin{eqnarray}
\left[ T_{1/2}^{\left( 0\nu \right) }\right] ^{-1} &=&\frac{\left\vert
\left\langle m_{\nu }\right\rangle \right\vert }{m_{e}}^{2}C_{mm}+\frac{%
\left\vert \left\langle m_{\nu }\right\rangle \right\vert }{m_{e}}%
\left\langle \lambda \right\rangle C_{m\lambda }+\frac{\left\vert
\left\langle m_{\nu }\right\rangle \right\vert }{m_{e}}\left\langle \eta
\right\rangle C_{m\eta }+\left\langle \lambda \right\rangle ^{2}C_{\lambda
\lambda }  \notag \\
&&+\left\langle \eta \right\rangle ^{2}C_{\eta \eta }+\left\langle \lambda
\right\rangle \left\langle \eta \right\rangle C_{\lambda \eta },
\end{eqnarray}

\noindent where 
\begin{eqnarray}
\left\langle m_{\nu }\right\rangle &=&\sum\nolimits_{i}^{\prime
}U_{ei}^{2}m_{i}, \\
\left\langle \lambda \right\rangle &=&\lambda \left\vert
\sum\nolimits_{i}^{^{\prime }}\left( \frac{g_{V}^{\prime }}{g_{V}}\right)
U_{ei}V_{ei}\right\vert , \\
\left\langle \eta \right\rangle &=&\eta \left\vert
\sum\nolimits_{i}^{^{\prime }}U_{ei}V_{ei}\right\vert ,
\end{eqnarray}%
\noindent\ and the nuclear structure factors $C_{xy}$ are written as 
%\begin{align}
\begin{subequations}
\begin{eqnarray}
C_{mm} &=&G_{01}\left\vert M^{\left( 0\nu \right) }\right\vert ^{2}, \\
C_{m\lambda } &=&M^{\left( 0\nu \right) }\left(
G_{04}M_{1+}-G_{03}M_{2-}\right) , \\
C_{m\eta } &=&M^{\left( 0\nu \right)
}(G_{03}M_{2+}-G_{04}M_{1-}-G_{05}M_{P}+G_{06}M_{R}), \\
C_{\lambda \lambda } &=&G_{02}\left\vert M_{2-}\right\vert ^{2}-\frac{2}{9}%
G_{03}(M_{1+}M_{2-})+\frac{1}{9}G_{04}\left\vert M_{1+}\right\vert ^{2}, \\
C_{\eta \eta } &=&G_{02}\left\vert M_{2+}\right\vert ^{2}-\frac{2}{9}%
G_{03}(M_{1-}M_{2+})+\frac{1}{9}G_{04}\left\vert M_{1-}\right\vert ^{2} 
\notag \\
&&-G_{07}(M_{P}M_{R})+G_{08}\left\vert M_{P}\right\vert
^{2}+G_{09}\left\vert M_{R}\right\vert ^{2}, \\
C_{\lambda \eta } &=&-2G_{02}(M_{2+}M_{2-})+\frac{2}{9}%
G_{03}(M_{2+}M_{1+}+M_{2-}M_{1-})  \notag \\
&&-\frac{2}{9}G_{04}(M_{1-}M_{1+}).
\end{eqnarray}
%\end{align}
In addition, the combinations of NTMEs $M^{\left( 0\nu \right) }$ and $%
M_{i\pm }$ ($i=1,2$) are defined as 
\end{subequations}
\begin{eqnarray}
M^{\left( 0\nu \right) } &=&M_{GT}-M_{F}+M_{T}, \\
M_{1\pm } &=&M_{qGT}-6M_{qT}\pm 3M_{qF}, \\
M_{2\pm } &=&M_{\omega GT}\pm M_{\omega F}-\frac{1}{9}M_{1\mp }.
\end{eqnarray}

Employing the generally agreed closure approximation in conjunction with the
HFB wave functions, the NTMEs $M_{\alpha }$ ($\alpha =F,GT,T,\omega F,\omega
GT,qF,qGT,qT,P$ and $R$) appearing in the expressions of nuclear structure
factors $C_{xy}$ are calculated by using the following expression \cite%
{rath10}. 
\begin{eqnarray}
M_{\alpha } &=&\left\langle 0_{f}^{+}\left\Vert O_{\alpha }(\mathbf{r},%
\mathbf{\sigma )}\right\Vert 0_{i}^{+}\right\rangle  \notag \\
&=&\left[ n^{J_{f}=0}n^{Ji=0}\right] ^{-1/2}\int\limits_{0}^{\pi }d\theta
sin\theta n_{(Z,N),(Z+2,N-2)}(\theta )  \notag \\
&&\times \sum\limits_{\alpha \beta \gamma \delta }\left\langle \alpha \beta
\left\vert O_{\alpha }(\mathbf{r},\mathbf{\sigma )}\right\vert \gamma \delta
\right\rangle \times \sum\limits_{\varepsilon \eta }\frac{\left(
f_{Z+2,N-2}^{(\pi )\ast }\right) _{\varepsilon \beta }}{\left[ \left(
1+F_{Z,N}^{(\pi )}(\theta )f_{Z+2,N-2}^{(\pi )\ast }\right) \right]
_{\varepsilon \alpha }}  \notag \\
&&\times \frac{\left( F_{Z,N}^{(\nu )\ast }\right) _{\eta \delta }}{\left[
\left( 1+F_{Z,N}^{(\nu )}(\theta )f_{Z+2,N-2}^{(\nu )\ast }\right) \right]
_{\gamma \eta }}.  \label{m0n}
\end{eqnarray}

The calculation of $n^{J}$, $n_{(Z,N),(Z+2,N-2)}(\theta )$, $f_{Z,N}$ and $%
F_{Z,N}(\theta )$ require the intrinsic wave functions ${|\Phi _{0}\rangle }$
of axially symmetric state with $K=0$ expressed by the amplitudes $%
(u_{im},v_{im})$ and expansion coefficients $C_{ij,m}$, which are in turn
obtained by minimizing the expectation value of the effective Hamiltonian
given by \cite{rath10} 
\begin{equation}
H=H_{sp}+V(P)+V(QQ)+V(HH),  \label{hami}
\end{equation}%
in a basis constructed by using a set of deformed states. In Eq. (\ref{hami}%
), the $H_{sp}$, $V(P)$, $V(QQ)$ and $V(HH)$ denote the single particle
Hamiltonian, the pairing, quadrupole-quadrupole and
hexadecapole-hexadecapole parts of the effective two-body interaction,
respectively. Further, the transition operators have the following general
structure 
\begin{equation}
O_{\alpha }\left( \mathbf{r},\mathbf{\sigma ,\tau }\right) =S_{\alpha
}\left( \mathbf{r},\mathbf{\sigma }\right) \tau _{n}^{+}\tau _{m}^{+}\frac{2R%
}{\pi }\int h_{\alpha }(qr)f_{\alpha }(q^{2})q^{2}dq.
\end{equation}%
The calculation of $M^{\left( 0\nu \right) }$ has already been discussed in
Ref. \citen{rath13}. Neglecting the induced pseudoscalar terms in the
nonrelativistic reduction of right-handed $V+A$ current \cite{stef15}, the
explicit structure of $S_{\alpha }\left( \mathbf{r},\mathbf{\sigma }\right) $%
, $h_{\alpha }(qr)$ and $f_{\alpha }(q^{2})$ for the rest of the NTMEs $%
M_{\alpha }$ is given in Table~1.

\begin{table}[pt]
\tbl{Explicit structure of $S_{\alpha }\left( \mathbf{r},\mathbf{\sigma }\right) $, $h_{\alpha }(qr)$ and $f_{\alpha }(q^{2})$ of transition operator 
$O_{\alpha }\left( \mathbf{r},\mathbf{\sigma ,\tau }\right)$.} {\begin{tabular}{lccl}
\toprule NTME & $S_{\alpha }\left( \mathbf{r},\mathbf{\sigma }\right) $ & $h_{\alpha }(qr)$ & $f_{\alpha }(q^{2})$ \\ 
\colrule $M_{\omega F}$ & $1$ & $\dfrac{j_{0}(qr)}{\left( q+\overline{A}\right) ^{2}}$ & $\dfrac{g_{V}^{2}(q^{2})}{g_{A}^{2}}$ \\ 
$M_{\omega GT}$ & $\sigma _{1}\cdot \sigma _{2}$ & $\dfrac{j_{0}(qr)}{\left(
q+\overline{A}\right) ^{2}}$ & $\dfrac{g_{A}^{2}(q^{2})}{g_{A}^{2}}$ \\ 
$M_{qF}$ & $1$ & $\dfrac{j_{1}(qr)qr}{q\left( q+\overline{A}\right) }$ & $\dfrac{g_{V}^{2}(q^{2})}{g_{A}^{2}}$ \\ 
$M_{qGT}$ & $\sigma _{1}\cdot \sigma _{2}$ & $\dfrac{j_{1}(qr)qr}{q\left( q+\overline{A}\right) }$ & $\dfrac{g_{A}^{2}(q^{2})}{g_{A}^{2}}$ \\ 
$M_{qT}$ & $3(\sigma _{1}\cdot \hat{\mathbf{r}}_{12})\left( \mathbf{\sigma }_{1}\cdot \hat{\mathbf{r}}_{12}\right) -\sigma _{1}\cdot \sigma _{2}$ & $\dfrac{j_{1}(qr)qr}{q\left( q+\overline{A}\right) }$ & $\dfrac{g_{A}^{2}(q^{2})}{3g_{A}^{2}}$ \\ 
$M_{P}$ & $i\dfrac{R}{2r^{2}}\left( \mathbf{\sigma }_{1}-\mathbf{\sigma }_{2}\right) \cdot \left( \dfrac{\mathbf{r}\times \mathbf{r}_{+}}{R}\right) $
& $\dfrac{j_{1}(qr)qr}{q\left( q+\overline{A}\right) }$ & $\dfrac{g_{A}(q^{2})g_{V}(q^{2})}{g_{A}^{2}}$ \\ 
$M_{R}$ & $\sigma _{1}\cdot \sigma _{2}$ & $\dfrac{j_{0}(qr)q^{2}}{q\left( q+\overline{A}\right) }$ & $\dfrac{1}{3m_{N}}\left( 1+\dfrac{g_{M}\left(
q^{2}\right) }{g_{V}(q^{2})}\right) \dfrac{g_{A}(q^{2})g_{V}(q^{2})}{g_{A}^{2}}$ \\ 
\botrule &  &  & 
\end{tabular}
}
\end{table}

\section{Results and Discussions}

Employing the PHFB approach, four different sets of wave functions were
generated with the consideration of four different parametrizations of the
two body effective interaction \cite{rath10}. The strength parameters of $%
V(QQ)$, namely proton-proton, neutron-neutron and proton-neutron components
are denoted by $\chi _{2pp}$, $\chi _{2nn}$ and $\chi _{2pn}$, respectively.
Two different parametrizations, denoted by \textit{PQQ1} and \textit{PQQ2}
were obtained by fitting the excitation energy $E_{2^{+}}$ of the $\ $2$^{+}$
state either by taking $\chi _{2pp}=\chi _{2nn}$ and varying the strength of 
$\chi _{2pn}$ or by taking $\chi _{2pp}=\chi _{2nn}=\chi _{2pn}/2$ and
varying the three parameters together. Two additional parametrizations,
namely \textit{PQQHH1} and \textit{PQQHH2} were obtained with the inclusion
of the hexadecapolar \textit{HH} part of the effective interaction.

By comparing the theoretically calculated yrast spectra, the reduced $B(E2$:$%
0^{+}\rightarrow 2^{+})$ transition probabilities, deformation parameters $%
\beta _{2}$, static quadrupole moments $Q(2^{+})$, gyromagnetic factors $%
g(2^{+})$ and NTMEs $M_{2\nu }$ for the $0^{+}\rightarrow 0^{+}$ transition
with the available experimental data, the reliability of the wave functions
had been ascertained in Ref. \citen{rath10}. Moreover, the same wave
functions had been employed for the study of $0\nu \beta ^{-}\beta ^{-}$
decay of $^{94,96}$Zr, $^{98,100}$Mo, $^{104}$Ru, $^{110}$Pd, $^{128,130}$Te
and $^{150}$Nd isotopes within mechanisms involving exchange of light as
well as heavy Majorona neutrinos, classical Majorons, sterile neutrinos \cite%
{rath12,rath13} and new Majorons \cite{rath16}. 
\begin{table}[pt]
\tbl{Change in the NTME $M_{\alpha }$ of $0\nu \beta ^{-}\beta ^{-}$decay (in \%) due to the exchange of light Majorana neutrino, and admixture
of $V-A$ and $V+A$ currents, with the inclusion of FNS and SRC (SRC1, SRC2,
and SRC3) for all four parametrizations of the effective two-body
interaction.} {\begin{tabular}{lccrrrrrr}
\toprule {\small NTME} &  & {\small FNS} &  & \multicolumn{5}{c}{\small FNS+SRC} \\ \cline{5-9}
& ~~~~ & \multicolumn{1}{r}{} & ~~~~ & {\small SRC1} & \multicolumn{1}{r}{}
& {\small SRC2} & ~~~~ & {\small SRC3} \\ 
\colrule ${\small M}_{\omega F}$ &  & \multicolumn{1}{r}{\small 13.1--17.7}
&  & {\small 11.6--17.3} &  & {\small 0.1--1.1} &  & {\small 3.1--3.8} \\ 
${\small M}_{qF}$ &  & \multicolumn{1}{r}{\small 25.8--37.7} &  & {\small 2.9--5.9} &  & {\small 3.5--5.3} &  & {\small 4.2--6.6} \\ 
${\small M}_{\omega GT}$ &  & \multicolumn{1}{r}{\small 9.0--11.2} &  & 
{\small 13.9--18.0} &  & {\small 1.3--2.5} &  & {\small 2.6--3.0} \\ 
${\small M}_{qGT}$ &  & \multicolumn{1}{r}{\small 18.9--24.2} &  & {\small 5.3--7.6} &  & {\small 3.1--4.0} &  & {\small 4.3--5.8} \\ 
${\small M}_{qT}$ &  & \multicolumn{1}{r}{\small 0.2--34.2} &  & {\small 0.0--2.4} &  & {\small 0.0--2.3} &  & {\small 0.0--2.0} \\ 
${\small M}_{P}$ &  & \multicolumn{1}{r}{\small 10.8--42.5} &  & {\small 2.9--17.8} &  & {\small 3.8--14.1} &  & {\small 4.5--17.9} \\ 
${\small M}_{R}$ &  & \multicolumn{1}{r}{\small 30.9--34.6} &  & {\small 55.2--56.6} &  & {\small 28.1--29.4} &  & {\small 10.4--11.2} \\ 
\botrule &  &  &  &  &  &  &  & 
\end{tabular}
}
\end{table}

In order to estimate average NTMEs $M_{\alpha }$ and uncertainties $\Delta
M_{\alpha }$ statistically, sets of twelve NTMEs are calculated by using Eq.
(\ref{m0n}) with the consideration of four different parametrizations of the
two body effective interaction \cite{rath10} and three different
parametrizations of the SRC \cite{simk09}. By considering a Jastrow form of
short range correlations, three different parametrizations of SRC have been
given by \cite{simk09} 
\begin{equation}
f(r)=1-ce^{-ar^{2}}(1-br^{2}),
\end{equation}%
where $a=1.1$ $fm^{-2}$, $1.59$ $fm^{-2}$, $1.52$ $fm^{-2}$, $b=0.68$ $%
fm^{-2}$, $1.45$ $fm^{-2}$, $1.88$ $fm^{-2}$ and $c=1.0$, $0.92$, $0.46$ for
Miller and Spencer parametrization, Argonne NN and CD-Bonn potentials, and
are denoted by SRC1, SRC2 and SRC3, respectively. Specifically, sets of
twelve NTMEs, namely $M_{\omega F,qF}$, $M_{\omega GT,qGT}$, $M_{qT}$, $%
M_{P} $, and $M_{R}$ are calculated within the approximations of point
nucleons (P), nucleons having finite size ((FNS) and also with SRC (FNS+SRC).

In Table~2, the relative changes in NTMEs $M_{\alpha }$ (in \%) due to the
different approximations are presented. Due to FNS, the maximum change in $%
{\small M}_{\omega F,\omega GT}$, ${\small M}_{qF,qGT,qT,P}$ and ${\small M}%
_{R}$ is about 18\%, 40\%, and 35\%, respectively. With the inclusion of
SRC, the NTMEs ${\small M}_{\omega F,\omega GT}$ change by about 18\%, 2.5\%
and 4\% due to SRC1, SRC2 and SRC3, respectively. The observed changes in $%
{\small M}_{qF,qGT}$ with the inclusion of SRC1, SRC2 and SRC3 are of the
same order and the maximum change is about 7\%. Due to the inclusion of SRC,
the change in ${\small M}_{qT}$ is about 2\% and ${\small M}_{P}$ can change
between 2\%-18\%. The maximum change in ${\small M}_{R}$ due to SRC1, SRC2
and SRC3 is about 57\%, 29\% and 11\%, respectively. To quantify the effect
of deformation on $M_{\alpha }$, the quantity $D_{\alpha }=$ $M_{\alpha
}(\zeta _{qq}=0)/M_{\alpha }(\zeta _{qq}=1)$ has been defined as the ratio
of $M_{\alpha }$ at zero deformation ($\zeta _{qq}=0$) and full deformation (%
$\zeta _{qq}=1$) \cite{chat08}. In the Table~3, we tabulate the values of $%
D_{\alpha }$ for $\alpha =\omega F,\omega GT,qF,qGT,qT,P$ and $R$. In the
mass range $A=90-150$, the NTMEs $M_{\alpha }$ are suppressed by factor of
about 2--7 ($D_{qT}$ and $D_{P}$ are suppressed by a factor of about 25 and
14, respectively) due to deformation effects and hence, a proper
consideration of deformation of participating nuclei is quite crucial in the
nuclear structure aspects of $0\nu \beta ^{-}\beta ^{-}$ decay. 
\begin{table}[pt]
\tbl{Deformation ratio $D_{\alpha }$ with the inclusion of FNS and 
SRC (SRC1, SRC2, and SRC3) for all four parametrizations of the effective 
two-body interaction.} {\begin{tabular}{lrrrrrr}
\toprule ${\small D}_{\alpha }$ &  & \multicolumn{5}{c}{\small FNS+SRC} \\ 
\cline{3-7}
& ~~~~ & {\small SRC1} & ~~~~ & {\small SRC2} & ~~~~ & {\small SRC3} \\ 
\colrule ${\small D}_{\omega F}$ &  & {\small 1.9--6.9} &  & {\small 1.8--6.9} &  & {\small 1.8--6.9} \\ 
${\small D}_{qF}$ &  & {\small 1.9--6.9} &  & {\small 1.9--6.9} &  & {\small 1.9--6.9} \\ 
${\small D}_{\omega GT}$ &  & {\small 1.9--7.2} &  & {\small 1.9--7.2} &  & 
{\small 1.9--7.2} \\ 
${\small D}_{qGT}$ &  & {\small 2.0--7.2} &  & {\small 1.9--7.1} &  & 
{\small 1.9--7.1} \\ 
${\small D}_{qT}$ &  & {\small -0.8--25.0} &  & {\small -0.8--24.9} &  & 
{\small -0.8--25.0} \\ 
${\small D}_{P}$ &  & {\small 1.8--13.6} &  & {\small 1.8--12.1} &  & 
{\small 1.8--11.9} \\ 
${\small D}_{R}$ &  & {\small 1.8--7.1} &  & {\small 1.8--7.1} &  & {\small 1.8--7.0} \\ 
\botrule &  &  &  &  &  & 
\end{tabular}
}
\end{table}

In Table~4, the averages and standard deviations of seven NTMEs,
namely $M_{\omega F,qF}$, $M_{\omega GT,qGT}$, $M_{qT}$, $M_{P}$ and $M_{R}$
are compared with those calculated employing QRPA \cite{muto89} and QRPA
with a partial restoration of isospin \cite{simk17}. It is observed that the
\begin{table}[tp]
\tbl{Average values for NTMEs $\overline{M}_{\alpha }$ (uncertainty $\Delta \overline{M}_{\alpha })$ for the $0\nu \beta ^{-}\beta ^{-}$ decay of 
$^{94,96}$Zr, $^{100}$Mo, $^{110}$Pd, $^{128,130}$Te and $^{150}$Nd isotopes.}
{%
\begin{tabular}{lrrrrrrrr}
\toprule NTMEs &  & $^{{\small 94}}${\small Zr} & $^{{\small 96}}${\small Zr}
& $^{{\small 100}}${\small Mo} & $^{{\small 110}}${\small Pd} & $^{{\small %
128}}${\small Te} & $^{{\small 130}}${\small Te} & $^{{\small 150}}${\small %
Nd} \\ 
\colrule $\overline{{\small M}}_{\omega F}$ &  & \multicolumn{1}{r}{\small %
0.569} & \multicolumn{1}{r}{\small 0.443} & \multicolumn{1}{r}{\small 1.004}
& \multicolumn{1}{r}{\small 1.102} & \multicolumn{1}{r}{\small 0.587} & 
\multicolumn{1}{r}{\small 0.642} & \multicolumn{1}{r}{\small 0.456} \\ 
$\Delta \overline{{\small M}}_{\omega F}$ &  & \multicolumn{1}{r}{\small %
0.066} & \multicolumn{1}{r}{\small 0.050} & \multicolumn{1}{r}{\small 0.130}
& \multicolumn{1}{r}{\small 0.150} & \multicolumn{1}{r}{\small 0.061} & 
\multicolumn{1}{r}{\small 0.081} & \multicolumn{1}{r}{\small 0.071} \\ 
$\overline{{\small M}}_{qF}$ &  & \multicolumn{1}{r}{\small 0.627} & 
\multicolumn{1}{r}{\small 0.470} & \multicolumn{1}{r}{\small 1.115} & 
\multicolumn{1}{r}{\small 1.259} & \multicolumn{1}{r}{\small 0.699} & 
\multicolumn{1}{r}{\small 0.779} & \multicolumn{1}{r}{\small 0.567} \\ 
$\Delta \overline{{\small M}}_{qF}$ &  & \multicolumn{1}{r}{\small 0.058} & 
\multicolumn{1}{r}{\small 0.055} & \multicolumn{1}{r}{\small 0.156} & 
\multicolumn{1}{r}{\small 0.185} & \multicolumn{1}{r}{\small 0.065} & 
\multicolumn{1}{r}{\small 0.114} & \multicolumn{1}{r}{\small 0.094} \\ 
$\overline{{\small M}}_{\omega GT}$ &  & \multicolumn{1}{r}{\small -3.119} & 
\multicolumn{1}{r}{\small -2.303} & \multicolumn{1}{r}{\small -4.985} & 
\multicolumn{1}{r}{\small -5.618} & \multicolumn{1}{r}{\small -2.849} & 
\multicolumn{1}{r}{\small -3.140} & \multicolumn{1}{r}{\small -2.134} \\ 
$\Delta \overline{{\small M}}_{\omega GT}$ &  & \multicolumn{1}{r}{\small %
0.312} & \multicolumn{1}{r}{\small 0.230} & \multicolumn{1}{r}{\small 0.516}
& \multicolumn{1}{r}{\small 0.596} & \multicolumn{1}{r}{\small 0.335} & 
\multicolumn{1}{r}{\small 0.360} & \multicolumn{1}{r}{\small 0.324} \\ 
$\overline{{\small M}}_{qGT}$ &  & \multicolumn{1}{r}{\small -3.841} & 
\multicolumn{1}{r}{\small -2.799} & \multicolumn{1}{r}{\small -6.081} & 
\multicolumn{1}{r}{\small -7.068} & \multicolumn{1}{r}{\small -3.541} & 
\multicolumn{1}{r}{\small -3.969} & \multicolumn{1}{r}{\small -2.819} \\ 
$\Delta \overline{{\small M}}_{qGT}$ &  & \multicolumn{1}{r}{\small 0.318} & 
\multicolumn{1}{r}{\small 0.183} & \multicolumn{1}{r}{\small 0.483} & 
\multicolumn{1}{r}{\small 0.591} & \multicolumn{1}{r}{\small 0.325} & 
\multicolumn{1}{r}{\small 0.455} & \multicolumn{1}{r}{\small 0.398} \\ 
$\overline{{\small M}}_{qT}$ &  & \multicolumn{1}{r}{\small 0.021} & 
\multicolumn{1}{r}{\small 0.050} & \multicolumn{1}{r}{\small 0.050} & 
\multicolumn{1}{r}{\small 0.065} & \multicolumn{1}{r}{\small 0.189} & 
\multicolumn{1}{r}{\small 0.084} & \multicolumn{1}{r}{\small 0.033} \\ 
$\Delta \overline{{\small M}}_{qT}$ &  & \multicolumn{1}{r}{\small 0.065} & 
\multicolumn{1}{r}{\small 0.024} & \multicolumn{1}{r}{\small 0.067} & 
\multicolumn{1}{r}{\small 0.073} & \multicolumn{1}{r}{\small 0.015} & 
\multicolumn{1}{r}{\small 0.005} & \multicolumn{1}{r}{\small 0.011} \\ 
$\overline{{\small M}}_{P}$ &  & \multicolumn{1}{r}{\small 2.382} & 
\multicolumn{1}{r}{\small 2.296} & \multicolumn{1}{r}{\small 3.966} & 
\multicolumn{1}{r}{\small 4.731} & \multicolumn{1}{r}{\small 1.091} & 
\multicolumn{1}{r}{\small 1.474} & \multicolumn{1}{r}{\small 0.260} \\ 
$\Delta \overline{{\small M}}_{P}$ &  & \multicolumn{1}{r}{\small 0.207} & 
\multicolumn{1}{r}{\small 0.121} & \multicolumn{1}{r}{\small 0.245} & 
\multicolumn{1}{r}{\small 0.241} & \multicolumn{1}{r}{\small 0.156} & 
\multicolumn{1}{r}{\small 0.073} & \multicolumn{1}{r}{\small 0.106} \\ 
$\overline{{\small M}}_{R}$ &  & \multicolumn{1}{r}{\small -2.274} & 
\multicolumn{1}{r}{\small -1.874} & \multicolumn{1}{r}{\small -3.832} & 
\multicolumn{1}{r}{\small -4.474} & \multicolumn{1}{r}{\small -2.541} & 
\multicolumn{1}{r}{\small -2.686} & \multicolumn{1}{r}{\small -1.801} \\ 
$\Delta \overline{{\small M}}_{R}$ &  & \multicolumn{1}{r}{\small 0.664} & 
\multicolumn{1}{r}{\small 0.542} & \multicolumn{1}{r}{\small 1.097} & 
\multicolumn{1}{r}{\small 1.279} & \multicolumn{1}{r}{\small 0.753} & 
\multicolumn{1}{r}{\small 0.758} & \multicolumn{1}{r}{\small 0.545} \\ \colrule
\multicolumn{9}{l}{{\small NTMEs with $g_A$=1.254 
in pnQRPA by (a) Muto }$et$ $al.$%
{\small \ \cite{muto89} and\ (b) \v{S}imkovic }$et${\small \ }$al.${\small \ 
\cite{simk17}}} \\ 
\colrule ${\small M}_{\omega F}$ & {\small (a)} &  &  & {\small -1.218} &  & 
{\small -1.047} & {\small -0.867} & {\small -1.630} \\ 
& {\small (b)} &  & {\small -1.117} & {\small -2.076} & {\small -2.015} &  & 
{\small -1.410} &  \\ 
${\small M}_{qF}$ & {\small (a)} &  &  & {\small -1.161} &  & {\small -1.054}
& {\small -0.860} & {\small -1.592} \\ 
& {\small (b)} &  & {\small -0.804} & {\small -1.588} & {\small -1.565} &  & 
{\small -0.995} &  \\ 
${\small M}_{\omega GT}$ & {\small (a)} &  &  & {\small 1.330} &  & {\small %
3.011} & {\small 2.442} & {\small 4.206} \\ 
& {\small (b)} &  & {\small 2.088} & {\small 4.159} & {\small 4.436} &  & 
{\small 3.091} &  \\ 
${\small M}_{qGT}$ & {\small (a)} &  &  & {\small -1.145} &  & {\small 1.999}
& {\small 1.526} & {\small 2.485} \\ 
& {\small (b)} &  & {\small 1.026} & {\small 2.389} & {\small 2.878} &  & 
{\small 1.746} &  \\ 
${\small M}_{qT}$ & {\small (a)} &  &  & {\small -0.823} &  & {\small -0.583}
& {\small -0.574} & {\small -1.148} \\ 
& {\small (b)} &  & {\small -0.200} & {\small -0.329} & {\small -0.281} &  & 
{\small -0.252} &  \\ 
${\small M}_{P}$ & {\small (a)} &  &  & {\small 1.182} &  & {\small -0.483}
& {\small -0.387} & {\small 0.998} \\ 
${\small M}_{R}$ & {\small (a)} &  &  & {\small 4.528} &  & {\small 4.371} & 
{\small 3.736} & {\small 7.005} \\ 
\botrule &  &  &  &  &  &  &  & 
\end{tabular}
}
\end{table}
\begin{table}[pt]
\tbl{Average nuclear structure factors $\overline{C}_{mm}$, $\overline{C}_{m\lambda }$, $\overline{C}_{m\eta }$, $\overline{C}_{\lambda \lambda }$, $\overline{C}_{\eta \eta }$ and $\overline{C}_{\lambda \eta }$ for the $0\nu
\beta ^{-}\beta ^{-}$ decay of $^{96}$Zr, $^{100}$Mo, $^{110}$Pd, $^{130}$Te
and $^{150}$Nd isotopes.} 
{\begin{tabular}{lccccc} \toprule
$\overline{C}_{\alpha \beta }$ & $^{{\small 96}}${\small Zr} & $^{%
{\small 100}}${\small Mo} & $^{{\small 110}}${\small Pd} & $^{{\small 130}}$%
{\small Te} & $^{{\small 150}}${\small Nd} \\ \colrule
$\overline{{\small C}}_{mm}$ & \multicolumn{1}{r}{{\small 4.37}$%
\times ${\small 10}$^{-13}$} & \multicolumn{1}{r}{{\small 1.62}$\times $%
{\small 10}$^{-12}$} & \multicolumn{1}{r}{{\small 6.42}$\times ${\small 10}$%
^{-13}$} & \multicolumn{1}{r}{{\small 6.09}$\times ${\small 10}$^{-13}$} &
\multicolumn{1}{r}{{\small 1.32}$\times ${\small 10}$^{-12}$} \\
$\overline{{\small C}}_{m\lambda }$ & \multicolumn{1}{r}{{\small -2.26}$%
\times ${\small 10}$^{-13}$} & \multicolumn{1}{r}{{\small -8.45}$\times $%
{\small 10}$^{-13}$} & \multicolumn{1}{r}{{\small -2.48}$\times ${\small 10}$%
^{-13}$} & \multicolumn{1}{r}{{\small -2.69}$\times ${\small 10}$^{-13}$} &
\multicolumn{1}{r}{{\small -6.85}$\times ${\small 10}$^{-13}$} \\
$\overline{{\small C}}_{m\eta }$ & \multicolumn{1}{r}{{\small 5.02}$\times $%
{\small 10}$^{-11}$} & \multicolumn{1}{r}{{\small 1.80}$\times ${\small 10}$%
^{-10}$} & \multicolumn{1}{r}{{\small 9.14}$\times ${\small 10}$^{-11}$} &
\multicolumn{1}{r}{{\small 6.97}$\times ${\small 10}$^{-11}$} &
\multicolumn{1}{r}{{\small 1.05}$\times ${\small 10}$^{-10}$} \\
$\overline{{\small C}}_{\lambda \lambda }$ & \multicolumn{1}{r}{{\small 1.51}%
$\times ${\small 10}$^{-12}$} & \multicolumn{1}{r}{{\small 4.76}$\times $%
{\small 10}$^{-12}$} & \multicolumn{1}{r}{{\small 8.21}$\times ${\small 10}$%
^{-13}$} & \multicolumn{1}{r}{{\small 1.21}$\times ${\small 10}$^{-12}$} &
\multicolumn{1}{r}{{\small 4.55}$\times ${\small 10}$^{-12}$} \\
$\overline{{\small C}}_{\eta \eta }$ & \multicolumn{1}{r}{{\small 1.15}$%
\times ${\small 10}$^{-8}$} & \multicolumn{1}{r}{{\small 3.52}$\times $%
{\small 10}$^{-8}$} & \multicolumn{1}{r}{{\small 1.45}$\times ${\small 10}$%
^{-8}$} & \multicolumn{1}{r}{{\small 1.19}$\times ${\small 10}$^{-8}$} &
\multicolumn{1}{r}{{\small 1.89}$\times ${\small 10}$^{-8}$} \\
$\overline{{\small C}}_{\lambda \eta }$ & \multicolumn{1}{r}{{\small -1.54}$%
\times ${\small 10}$^{-12}$} & \multicolumn{1}{r}{{\small -4.63}$\times $%
{\small 10}$^{-12}$} & \multicolumn{1}{r}{{\small -7.76}$\times ${\small 10}$%
^{-13}$} & \multicolumn{1}{r}{{\small -1.10}$\times ${\small 10}$^{-12}$} &
\multicolumn{1}{r}{{\small -4.08}$\times ${\small 10}$^{-12}$} \\ \botrule
\end{tabular}
}
\end{table}
maximum uncertainty in $\overline{M}_{\omega F,qF}$, $\overline{M}_{\omega
GT,qGT}$ and $\overline{M}_{P}$ is about 15\% but for $^{150}$Nd, in which
the standard deviation of $\overline{M}_{P}$ is about 40\%. In $^{94}$Zr, $%
^{100}$Mo, and $^{110}$Pd isotopes, the NTMEs $M_{qT}$ are quite uncertain
due to change of sign in the case of \textit{PQQHH1}, \textit{PQQHH2}, and 
\textit{PQQ2} parametrizations. The maximum uncertainty in $\overline{M}_{R}$
is about 30\%. In Ref. \citen{rath13}, NTMEs $M^{\left( 0\nu \right) }$ have
already been calculated. Presently, we reevaluate them for $g_{A}=1.2701$
and sets of twelve nuclear structure factors $C_{mm}$, $C_{m\lambda }$, $%
C_{m\eta }$, $C_{\lambda \lambda }$, $C_{\eta \eta }$ and $C_{\lambda \eta }$
are computed for $^{96}$Zr, $^{100}$Mo, $^{110}$Pd, $^{130}$Te and $^{150}$%
Nd isotopes using the phase space factors calculated by \v{S}tef\'{a}nik $et$
$al.$ \cite{stef15}. The averages of these six nuclear structure factors are
reported in Table~5. To exhibit the relative role of NTMEs due to different
mechanisms, we define $M_{eff}^{\left( 0\lambda \right) }$ and $%
M_{eff}^{\left( 0\eta \right) }$ as%
\begin{eqnarray}
C_{\lambda \lambda } &=&G_{01}\left\vert M_{eff}^{\left( 0\lambda \right)
}\right\vert ^{2}, \\
C_{\eta \eta } &=&G_{01}\left\vert M_{eff}^{\left( 0\eta \right)
}\right\vert ^{2},
\end{eqnarray}%
and present the NTMEs $M_{eff}^{\left( 0\lambda \right) }$ and $%
M_{eff}^{\left( 0\eta \right) }$ along with$\ M^{\left( 0\nu \right) }$ \cite%
{rath13} reevaluated for $g_{A}=1.2701$ in Table~6. It is observed that
\begin{table}[tp]
\tbl{Effective NTMEs $M_{eff}^{\left( 0\lambda \right) }$ and $M_{eff}^{\left( 0\eta \right) }$ along with $M^{\left( 0\nu \right) }$ for
the $0\nu \beta ^{-}\beta ^{-}$ decay of $^{96}$Zr, $^{100}$Mo, $^{110}$Pd, $^{130}$Te and $^{150}$Nd isotopes.}
{%
\begin{tabular}{clllllr}
\toprule {\small Nuclei} & ~~~~~~ & ${\small M}^{\left( 0\nu \right) }$ & 
~~~~~~~ & ${\small M}_{eff}^{\left( 0\lambda \right) }$ & ~~~~~~~ & ${\small %
M}_{eff}^{\left( 0\eta\right) }$ \\ 
\colrule $^{96}${\small Zr} &  & {\small 2.85} &  & \multicolumn{1}{c}%
{\small 5.30} &  & \multicolumn{1}{r}{\small 463.07} \\ 
$^{100}${\small Mo} &  & {\small 6.25} &  & \multicolumn{1}{c}{\small 10.71}
&  & \multicolumn{1}{r}{\small 920.40} \\ 
$^{110}${\small Pd} &  & {\small 7.15} &  & \multicolumn{1}{c}{\small 8.08}
&  & \multicolumn{1}{r}{\small 1072.80} \\ 
$^{130}${\small Te} &  & {\small 4.05} &  & \multicolumn{1}{c}{\small 5.71}
&  & \multicolumn{1}{r}{\small 567.26} \\ 
$^{150}${\small Nd} &  & {\small 2.84} &  & \multicolumn{1}{c}{\small 5.26}
&  & \multicolumn{1}{r}{\small 338.85} \\ 
\botrule &  &  &  &  &  & 
\end{tabular}
}
\end{table}
NTMEs $M_{eff}^{\left( 0\lambda \right) }$ are about twice of $M^{\left(
0\nu \right) }$ and NTMEs $M_{eff}^{\left( 0\eta \right) }$ are larger by
two orders in magnitude than the latter. 

The role of $\lambda $-mechanism of 0$\nu \beta ^{-}\beta ^{-}$ decay within
LRSM has recently been investigated in detail \cite{simk17}. Remarkably, a
number of observations, namely the near equality of $M_{v}=M^{(0\nu )}$ and $%
M_{2-}$, association of a single phase space factor $G_{02}$ with $\lambda $%
-mechanism, distinguishability of standard light Majorana neutrino mass and $%
\lambda $ mechanisms and exclusion of $\lambda $-mechanism as the dominant
mechanism of 0$\nu \beta ^{-}\beta ^{-}$ decay, have been reported.
Presently, we investigate the former three conclusions within PHFB approach.
In Fig. 1, the NTMEs $M_{v}=M^{(0\nu )}$, $M_{2-}$ and $M_{1+}$ for 
$^{96}$Zr, $^{100}$Mo, $^{110}$Pd, $^{130}$Te and $^{150}$Nd isotopes are
plotted. The near equality of $M_{v}$ and $M_{2-}$ as well as the
subdominant role of $M_{1+}/9$ in $M_{2-}$ can be easily noticed. The
ratios $f_{\lambda m}=C_{\lambda \lambda }/C_{mm}$ and \ $f_{\lambda
m}^{G}=G_{02}/G_{01}$ vs. $Q_{\beta \beta }$ are plotted in Fig. 2 and the
\begin{figure}[th]
\centerline{\includegraphics[width=10.3cm]{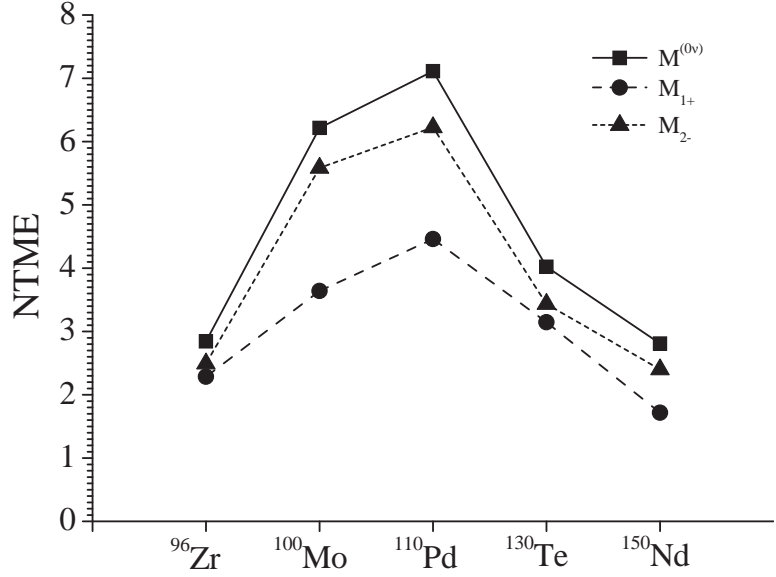}}
\caption{Comparison of NTMEs $M_v=M^{(0\nu)}, M_{1+}$
and $M_{2-}$.}
\end{figure}
\begin{figure}[th]
\centerline{\includegraphics[width=10.3cm]{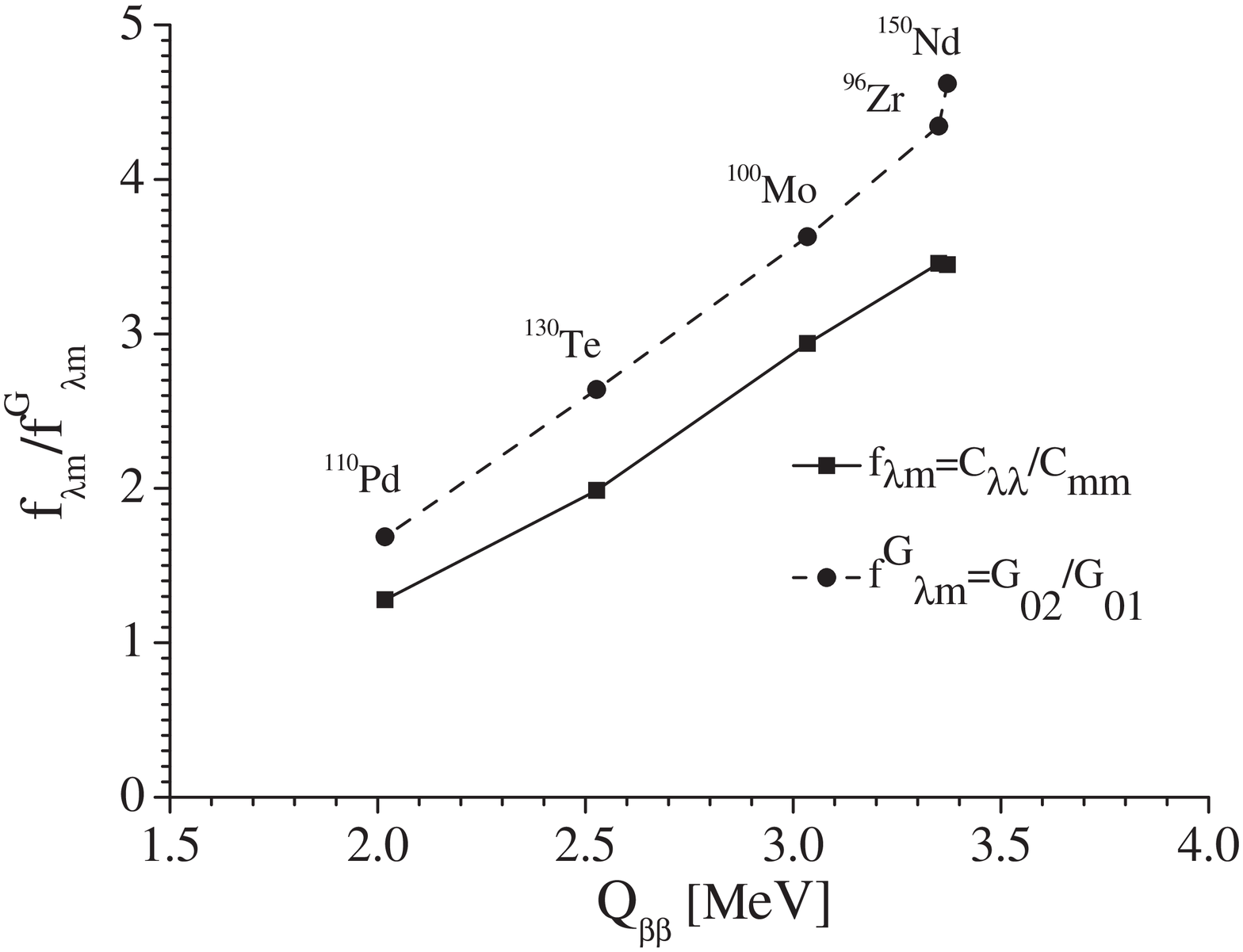}}
\caption{Ratios $f_{\lambda m}$ and $f_{\lambda m}^G$ 
as a function of $Q_{\beta\beta}$ \cite{stef15}.}
\end{figure}
association of a single phase space factor $G_{02}$ with $\lambda $%
-mechanism can be easily inferred.

Using the average nuclear structure factors $\overline{C}_{mm}$, $\overline{C%
}_{\lambda \lambda }$, $\overline{C}_{\eta \eta }$, on-axis limits on the
effective mass of light neutrino $\left\langle m_{\nu }\right\rangle $, the
effective weak coupling of right-handed leptonic current with right-handed
hadronic current $<\lambda >$, and the effective weak coupling of
right-handed leptonic current with left-handed hadronic current $<\eta >$
are extracted from the largest observed limits on half-lives $T_{1/2}^{(0\nu
)}$ of $0\nu \beta ^{-}\beta ^{-}$decay (Table~7). The extracted limits on $%
\left\langle m_{\nu }\right\rangle $, $\left\langle \lambda \right\rangle $,
and $\left\langle \eta \right\rangle $ for $^{130}$Te ($^{100}$Mo) nuclei
are $0.33$ eV ($0.38$ eV), $4.57\times 10^{-7}$ ($4.39\times 10^{-7}$) and $%
4.72\times 10^{-9}$ ($5.23\times 10^{-9}$), respectively. In the last two
columns of the same Table~7, the predicted half-lives $T_{1/2}^{(0\nu )}$ of 
$0\nu \beta ^{-}\beta ^{-}$decay of $^{96}$Zr, $^{100}$Mo, $^{110}$Pd, $%
^{130}$Te and $^{150}$Nd isotopes are given for two sets of parameters (i) $%
\left\langle m_{\nu }\right\rangle =50$ meV and (ii) $\left\langle m_{\nu
}\right\rangle =50$ meV, $\left\langle \lambda \right\rangle =10^{-7}$ and $%
\left\langle \eta \right\rangle =10^{-9}$. It is noticed that the predicted
half-lives $T_{1/2}^{(0\nu )}$ are smaller for the latter parametrization
than those of pure mass mechanism. By defining $\left[ T_{1/2}^{\left( 0\nu
\right) }\right] ^{-1}=C^{(0\nu )}$, it is seen that in total $C^{(0\nu )}$,
the the contribution of mass mechanism is about 13\%--17\%, the $\lambda $%
-term 
contributes 23\%--57\% and the $\eta $-term contributes 24\%-41\%. Further,
the contributions of $m\lambda $ and $m\eta $-term are about 7\%-8\% and
13\%--25\%, respectively, while the $\lambda \eta $-term contribute less
than 1\%. In Fig. 3, predicted half-lives $T_{1/2}^{(0\nu )}$ of $0\nu
\beta ^{-}\beta ^{-}$decay of $^{96}$Zr, $^{100}$Mo, $^{110}$Pd, $^{130}$Te
\begin{figure}[th]
\centerline{\includegraphics[width=10.3cm]{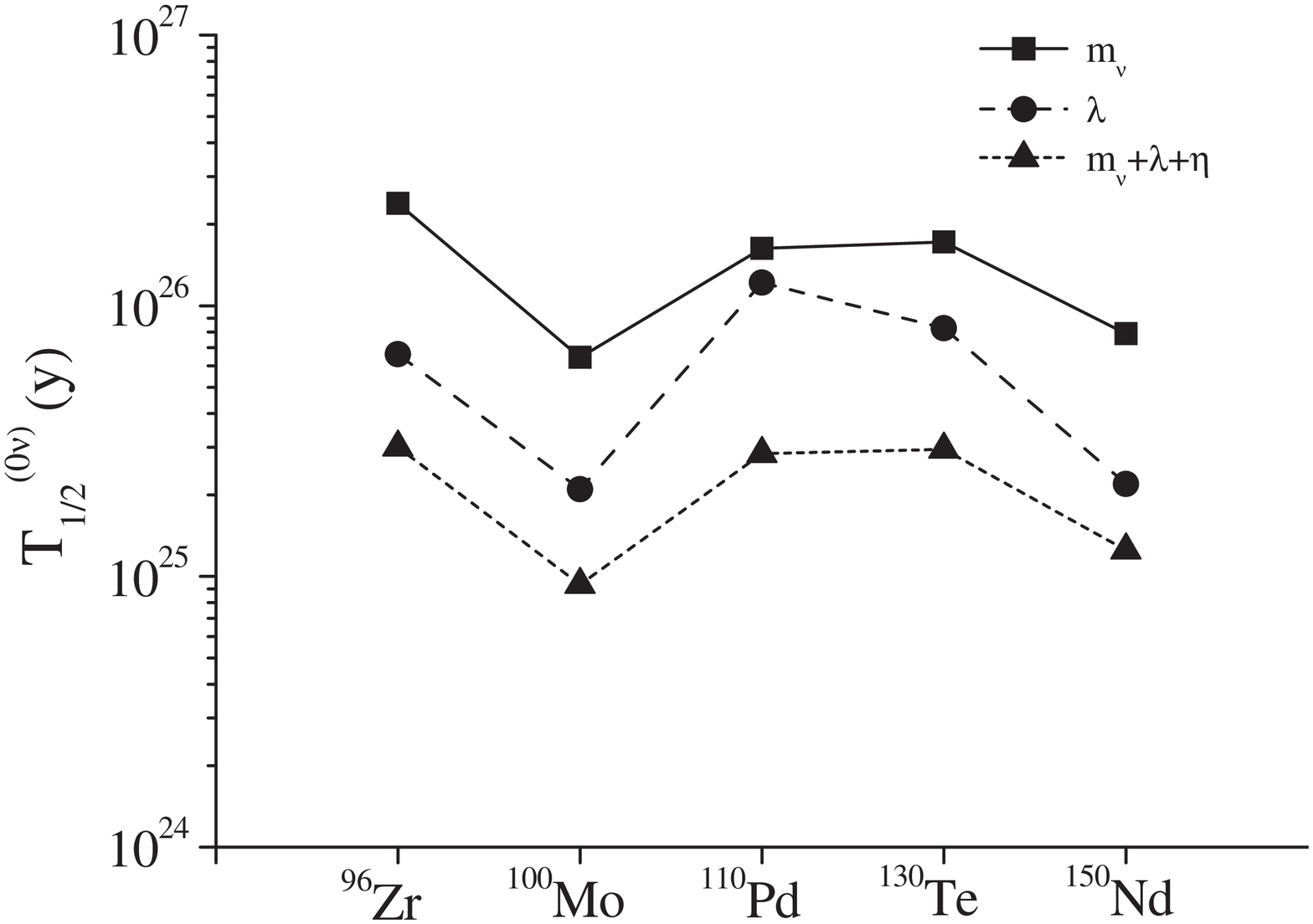}}
\caption{The half-lives $T_{1/2}^{(0\nu)}$ (y) in mass mechanism, $\lambda$ 
mechanism and mechanism involving both left and right handed currents.}
\end{figure}
and $^{150}$Nd isotopes for three sets of parameters (i) 
$\left\langle m_{\nu }\right\rangle =50$ meV, (ii) $\left\langle \lambda
\right\rangle =10^{-7}$ and (iii) $\left\langle m_{\nu }\right\rangle =50$
meV, $\left\langle \lambda \right\rangle =10^{-7}$, $\left\langle \eta
\right\rangle =10^{-9}$ are plotted. The distinguishability of $\lambda $%
-mechanism as well as mechanism involving both left and right handed
currents ($\eta $-term) in case of $0\nu \beta ^{-}\beta ^{-}$decay of $^{96}
$Zr, $^{100}$Mo, $^{110}$Pd, $^{130}$Te and $^{150}$Nd isotopes is
clearly exhibited. 
\begin{table}[tp]
\tbl{Experimental limits on half-lives $T_{1/2}^{(0\nu )}$ (y)
for the $0\nu \beta ^{-}\beta ^{-}$
decay of $^{96}$Zr \cite{argy10}, $^{100}$Mo \cite{arno15}, 
$^{110}$Pd \cite{wint52}, $^{130}$Te \cite{aldu16} and $^{150}$Nd \cite{arno16}
isotopes along with the extracted on-axis limits on the effective mass of 
light neutrino $\left\langle m_{\nu }\right\rangle $, 
$\left\langle \lambda \right\rangle $
and $\left\langle \eta \right\rangle $. 
Predicted half-lives $T_{1/2}^{(0\nu )}$ (y) of $0\nu \beta ^{-}\beta
^{-}$decay for two sets of parameters (i) $\left\langle m_{\nu
}\right\rangle =50$ meV (Case I) and (ii) $\left\langle m_{\nu
}\right\rangle =50$ meV, $\left\langle \lambda \right\rangle =10^{-7}$ and $\left\langle \eta \right\rangle =10^{-9}$ (Case II).}
{%
\begin{tabular}{clcccll}
\toprule {\small Nuclei} & $T_{1/2}^{(0\nu )}${\small (Exp)} & ${\small <m}%
_{\nu }{\small >}$ & ${\small <\lambda >}$ & ${\small <\eta >}$ & $%
T_{1/2}^{(0\nu )} ${\small (I)} & $T_{1/2}^{(0\nu )}${\small (II)} \\ 
\colrule $^{{\small 96}}${\small Zr} & {\small 9.2}$\times 10^{21}$ & 
{\small 8.09} & {\small 8.53}$\times ${\small 10}$^{-6}$ & {\small 1.00}$%
\times ${\small 10}$^{-7}$ & {\small 2.39}$\times ${\small 10}$^{26}$ & 
{\small 3.00}$\times ${\small 10}$^{25}$ \\ 
$^{{\small 100}}${\small Mo} & {\small 1.1}$\times 10^{24}$ & {\small 0.38}
& {\small 4.39}$\times ${\small 10}$^{-7}$ & {\small 5.23}$\times ${\small 10%
}$^{-9}$ & {\small 6.45}$\times ${\small 10}$^{25}$ & {\small 9.34}$\times $%
{\small 10}$^{24}$ \\ 
$^{{\small 110}}${\small Pd} & {\small 6.0}$\times 10^{17}$ & {\small 827} & 
{\small 1.43}$\times ${\small 10}$^{-3}$ & {\small 1.10}$\times ${\small 10}$%
^{-5}$ & {\small 1.63}$\times ${\small 10}$^{26}$ & {\small 2.84}$\times $%
{\small 10}$^{25}$ \\ 
$^{{\small 130}}${\small Te} & {\small 4.0}$\times 10^{24}$ & {\small 0.33}
& {\small 4.57}$\times ${\small 10}$^{-7}$ & {\small 4.72}$\times ${\small 10%
}$^{-9}$ & {\small 1.72}$\times ${\small 10}$^{26}$ & {\small 2.95}$\times $%
{\small 10}$^{25}$ \\ 
$^{{\small 150}}${\small Nd} & {\small 2.0}$\times 10^{22}$ & {\small 3.17}
& {\small 3.35}$\times ${\small 10}$^{-6}$ & {\small 5.36}$\times ${\small 10%
}$^{-8}$ & {\small 7.88}$\times ${\small 10}$^{25}$ & {\small 1.25}$\times $%
{\small 10}$^{25}$ \\ 
\botrule &  &  &  &  &  & 
\end{tabular}
}
\end{table}

\section{Conclusions}

Using HFB\ wave functions generated with four different parametrization of
pairing plus multipolar type of effective two body interaction, and three
different parametrizations of Jastrow SRC, sets of twelve NTMEs, namely $%
M_{\omega F,qF}$, $M_{\omega GT,qGT}$, $M_{qT}$, $M_{P}$, and $M_{R}$ are
calculated to study the $0\nu \beta ^{-}\beta ^{-}$ decay of $^{94,96}$Zr, $%
^{100}$Mo, $^{110}$Pd, $^{128,130}$Te and $^{150}$Nd isotopes within
mechanisms involving the light Majorana neutrino, and right handed $V+A$
current. The effect due to FNS is maximum (about 40\%) for ${\small M}%
_{qF,qGT,qT,P}$. Due to SRC1, SRC2 and SRC3, the maximum change in ${\small M%
}_{R}$ is about 57\%, 29\% and 11\%, respectively. Effects due to
deformation reduce the NTMEs by a factor of 2--7.

The maximum uncertainty in $M_{\omega F,qF}$, $M_{\omega GT,qGT}$ and $M_{P}$
is about 15\% albeit the standard deviation of $M_{P}$ for $^{150}$Nd is
about 40\%. In the case of $M_{R}$, the maximum uncertainty is about 30\%.
The NTMEs $M_{qT}$ are quite uncertain. Using the average nuclear structure
factors $\overline{C}_{mm}$, $\overline{C}_{\lambda \lambda }$, and $%
\overline{C}_{\eta \eta }$, the most stringent on-axis extracted limits on $%
\left\langle m_{\nu }\right\rangle $, $\left\langle \lambda \right\rangle $,
and $\left\langle \eta \right\rangle $ from the largest observed limits on
half-lives $T_{1/2}^{0\nu }$ of $^{130}$Te isotope are $0.33$ eV, $%
4.57\times 10^{-7}$ and $4.72\times 10^{-9}$, respectively. In the light of
Ref. \citen{simk17}, the role of $\lambda $-mechanism in the 0$\nu \beta
^{-}\beta ^{-}$ decay of $^{96}$Zr, $^{100}$Mo, $^{110}$Pd, $^{130}$Te and $%
^{150}$Nd isotopes is analyzed.

\section*{Acknowledgments}

This work is partially supported by DST-SERB, India vide sanction No.
SR/FTP/PS-085/2011, SB/S2/HEP-007/2013 and Council of Scientific and
Industrial Research (CSIR), India vide sanction No. 03(1216)/12/EMR-II.

%\begin{thebibliography}{000} %for 3 digits
%\begin{thebibliography}{00}  %for 2 digits

\end{document}